\documentclass[twocolumn]{revtex4-2}

\usepackage{graphicx} 
\usepackage{xcolor}
\usepackage{amsmath,amssymb,amsfonts,amsthm}
\begin{document}

\title{Improving propulsive efficiency using bio-inspired intermittent locomotion }
\author{Tristan Aur\'egan$^{1, 2}$}
\author{Mathilde Lemoine$^1$}
\author{Benjamin Thiria$^1$}
\author{Sylvain Courrech du Pont$^2$}

\affiliation{$^1$ Laboratoire de Physique et Mécanique des Milieux Hétérogènes, PMMH UMR 7636 CNRS
ESPCI PSL Research University, Université Paris Cité, Sorbonne Université, Paris, France.}
\affiliation{$^2$ Laboratoire Matière et Systèmes Complexes, UMR CNRS 7057, Université Paris Cité, Paris, France.}
\date{\today}

\maketitle

\section*{Abstract}
Many swimmers, especially small to medium-sized animals, use intermittent locomotion that differs from continuous swimming of large species. This type of locomotion, called burst-and-coast, is often associated with an energetic advantage. In this work, we investigate the intermittent locomotion inspired by fish locomotion but applied to a propeller. The energy consumption of burst-and-coast cycles is measured and compared to the continuous rotation regime. We show that a substantial drag ratio between the active and passive phases of the motion, as observed in fish, is critical for energy savings. Such a contrast can be obtained using a folding propeller that passively opens and closes as the propeller starts and stops rotating. For this reconfigurable propeller, intermittent propulsion is found to be energetically advantageous, saving up to 24\% of the energy required to cruise at a given speed. Using an analytical model, we show that intermittent motion is more efficient than continuous motion when the drag reduction in the coast phase exceeds 65\%. For fish-like locomotion, this threshold seems to be closer to 30\%. A formal analogy allows us to explain the difference between propeller propulsion and fish locomotion.

\section{Introduction}

Intermittent locomotion is widespread in nature, with many species using it to navigate their environment \cite{kramer_behavioral_2001}. Since drag increases quadratically with speed, this type of intermittent locomotion seems energetically unfavorable at first glance. 
However, a large number of flying or swimming animal species appear to use this type of locomotion precisely for the purpose of saving energy. Birds use two different variants of an intermittent flight style, either gliding or following a ballistic trajectory (bounding) \cite{tobalske_hovering_2010,rayner_aerodynamics_2001}, many fish species coast between each stroke \cite{wu_kinematics_2007,ribak_submerged_2005,xia_energy-saving_2018,coughlin_intermittent_2022,videler_swimming_1981}, just as humans have an energetic interest in gliding efficiently in breaststroke \cite{naemi_hydrodynamic_2010,barbosa_energetics_2010}. In addition, rowing crews on boats can optimize their performance by adjusting the frequency of their movements to minimize drag \cite{dode_wave_2022,brouwer_dont_2013}.

Weihs \cite{weihs_energetic_1974,weihs_energetic_1981}  was the first to theoretically investigate the relationship between efficiency and intermittent swimming cycles. He found that the alternation between active and passive phases (also called "burst and coast") can be energetically interesting for fish to swim at a given mean speed, depending on the kinematic parameters, {\it e.g.}, the duty cycle, which is the ratio of burst duration to coast duration, and is conditioned by the reduction in the drag experienced by the fish during the passive phase compared to the active phase. This reduction is due to a combination of a geometric effect (the body is aligned with the flow during the coast phase) and a dynamic effect (the tail movement increases the skin friction during the burst phase) \cite{ehrenstein_skin_2014,godoy-diana_diverse_2018}. 
From flow visualisation in experiments with koi carp, it has been estimated that the drag during the undulatory swimming (burst) is up to four times larger than when the fish's body is aligned with the flow (coast) \cite{wu_kinematics_2007,videler_swimming_1981}. This corresponds to a gain of 45\% of the energy expended at a given speed compared to continuous swimming. Recently, direct numerical flow simulations based on measurements of tetrafish swimming have shown that the set of parameters chosen by the fish (period, duty cycle, amplitude) corresponds to an optimum that minimizes energy expenditure \cite{li_burst-and-coast_2021,li_intermittent_2023}. The associated energy savings range from 3 to 26\% depending on the mean velocity.
\begin{figure*}[t]
    \centering
    \begin{minipage}{0.62\textwidth}
        \includegraphics[width=\textwidth]{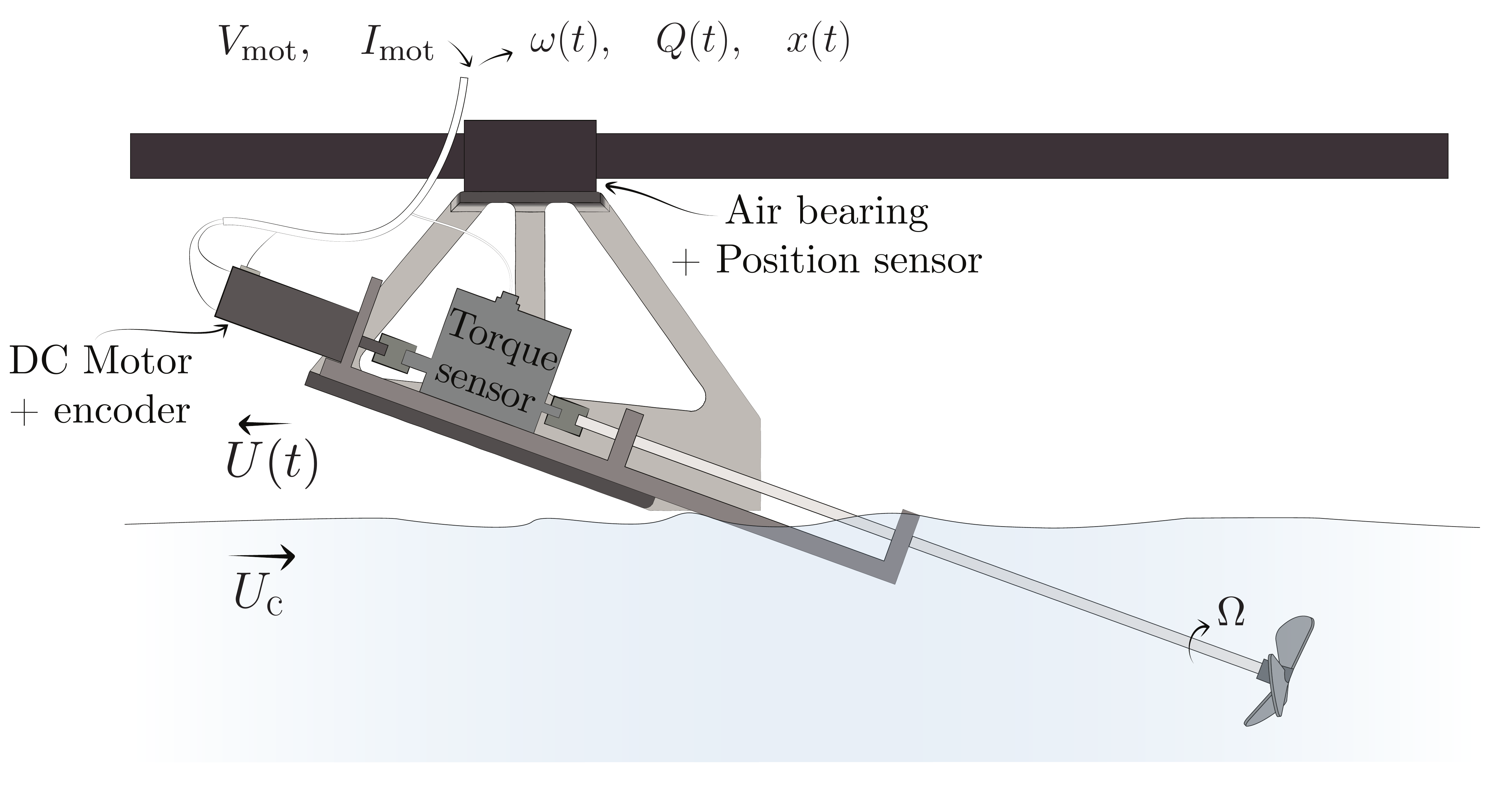}
      \end{minipage}
      \begin{minipage}{0.37\textwidth}
        \includegraphics[width = \textwidth]{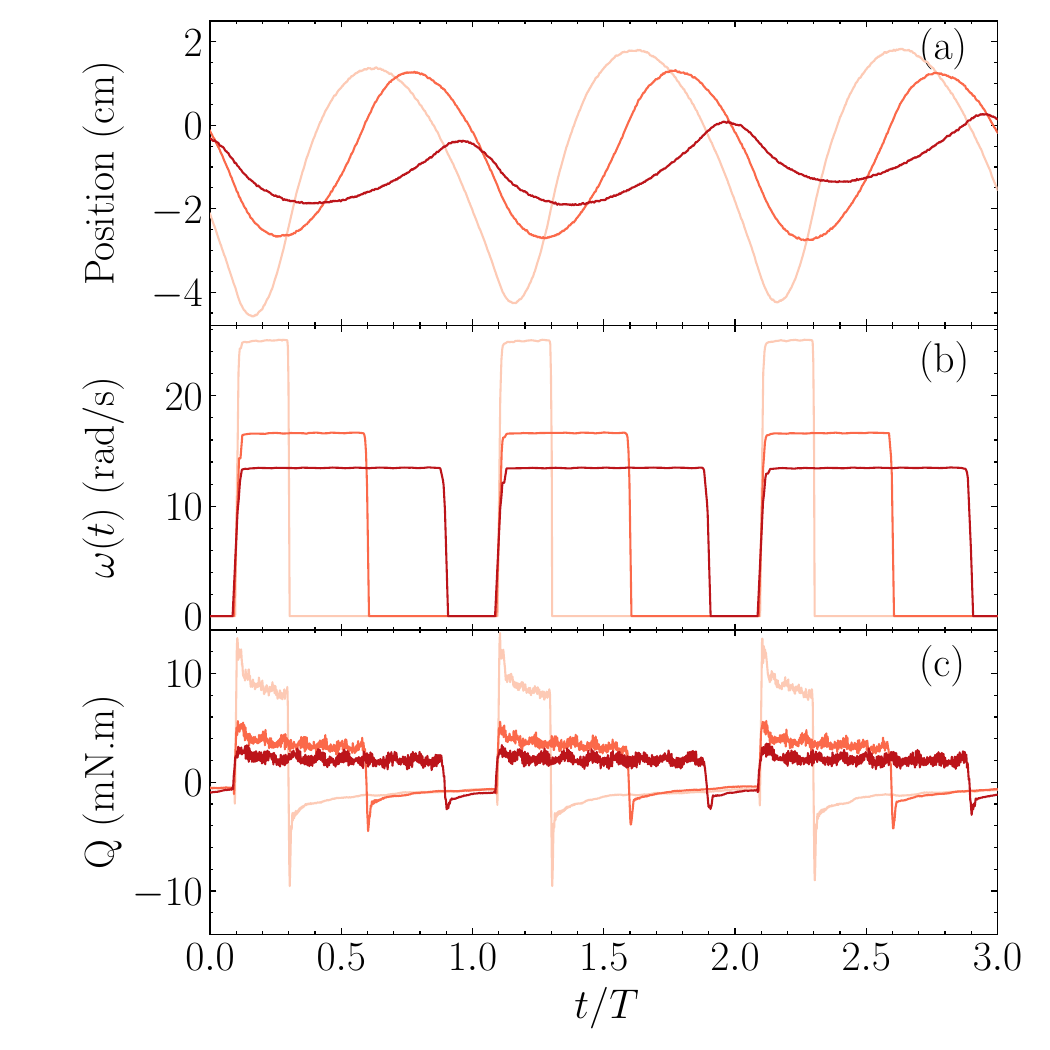}
      \end{minipage}
    \caption{(left) Schematic of the experimental setup, see Materials and Methods for details. (right) Examples of raw signals from experiments at the same cruising velocity and period ($U_\text{c} = 0.15$ m/s, $T = 5$ s, rigid propeller) for several duty cycles (0.2, 0.5, and 0.8, darker is closer to 1). (a) Position in the channel, (b) Angular velocity, (c) torque.}
    \label{fig:setup}
\end{figure*}
Several authors have tried experimentally or numerically to reproduce intermittent swimming using pitching or heaving airfoils or a combination of both \cite{floryan_forces_2017,akoz_unsteady_2018,akoz_intermittent_2021,akoz_large-amplitude_2019,gupta_body-caudal_2021}. Using a fixed setup in an imposed flow, Florian {\it et al.} \cite{floryan_forces_2017} showed that intermittent pitching  could be up to 20 \% more efficient than continuous actuation, and proposed a scaling law for cycle efficiency as a function of the duty cycle.

We investigate the energetic advantages of ``burst and coast" locomotion of a model-scale boat with a reconfigurable propeller that is intermittently actuated. The flow drives a passive reconfiguration of the propeller, which unfolds when rotation starts and folds when rotation stops.
Importantly, the boat can move freely back and forth in a channel with an imposed flow, integrating the non-stationary effects of intermittent locomotion that are detrimental to propulsion efficiency.
We find that intermittent propulsion can outperform continuous propulsion in reducing the total energy consumption by 24\%. We explain this significant energy reduction and the underlying physical mechanisms with an analytical model, and compare propeller propulsion and fish locomotion.

\section{Experimental results}\label{sec:results}

We performed burst and coast experiments using a self-propelled toy model scale boat that is free to move back and forth in a free-surface water channel with a controlled imposed flow at velocity $U_\text{c}$ (Fig. \ref{fig:setup}). The boat is mounted on a linear air bearing above the channel to ensure straight motion with negligible friction.
The propellers are 3D printed; they have three blades with a 30 degree pitch angle. The blades of the reconfigurable propeller are hinged and fold downstream when rotation stops. The motor is not disengaged during the coast phase, but is stopped.
Figure 2 shows the steady-state blade opening angle of the reconfigurable propeller as a function of the tip speed ratio, $\lambda = R\Omega / U_\text{c}$, when the propeller of radius $R$ is rotating at the angular velocity $\Omega$. The propeller is fully open when the tip speed ratio is greater than 4. In the burst and coast experiments, the tip speed ratio is always greater than this threshold when the propeller is rotating. The opening of the blades is driven by the hydrodynamic lift force.
With this folding propeller, the drag of the entire system is reduced by 80\% between the burst and coast phases, a value similar to that found by Wu for fish swimming \cite{wu_kinematics_2007}.

\begin{figure}[h!]
   \centering
   \includegraphics[width = \linewidth]{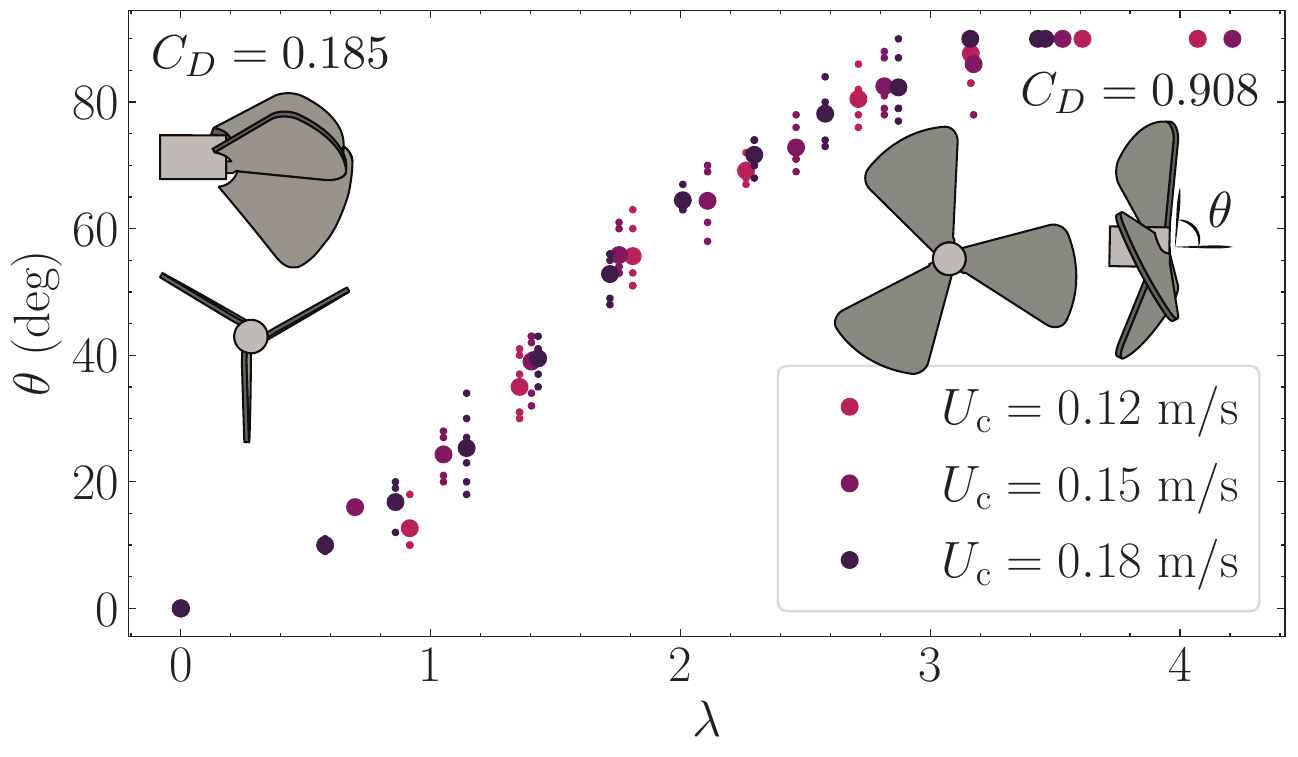}
   \caption{Folding propeller opening angle as a function of the tip speed ratio for several flow speeds. Smaller dots represent individual measurements, while larger ones represent the mean. The drag coefficient when closed is obtained via a direct measurement, while the one when opened is inferred from a fit of thrust produced by the rotor (see Appendix \ref{app:forces}).}
   \label{fig:opening}
\end{figure}

\begin{figure*}[htbp]
    \centering
    \includegraphics[width = 0.95\linewidth]{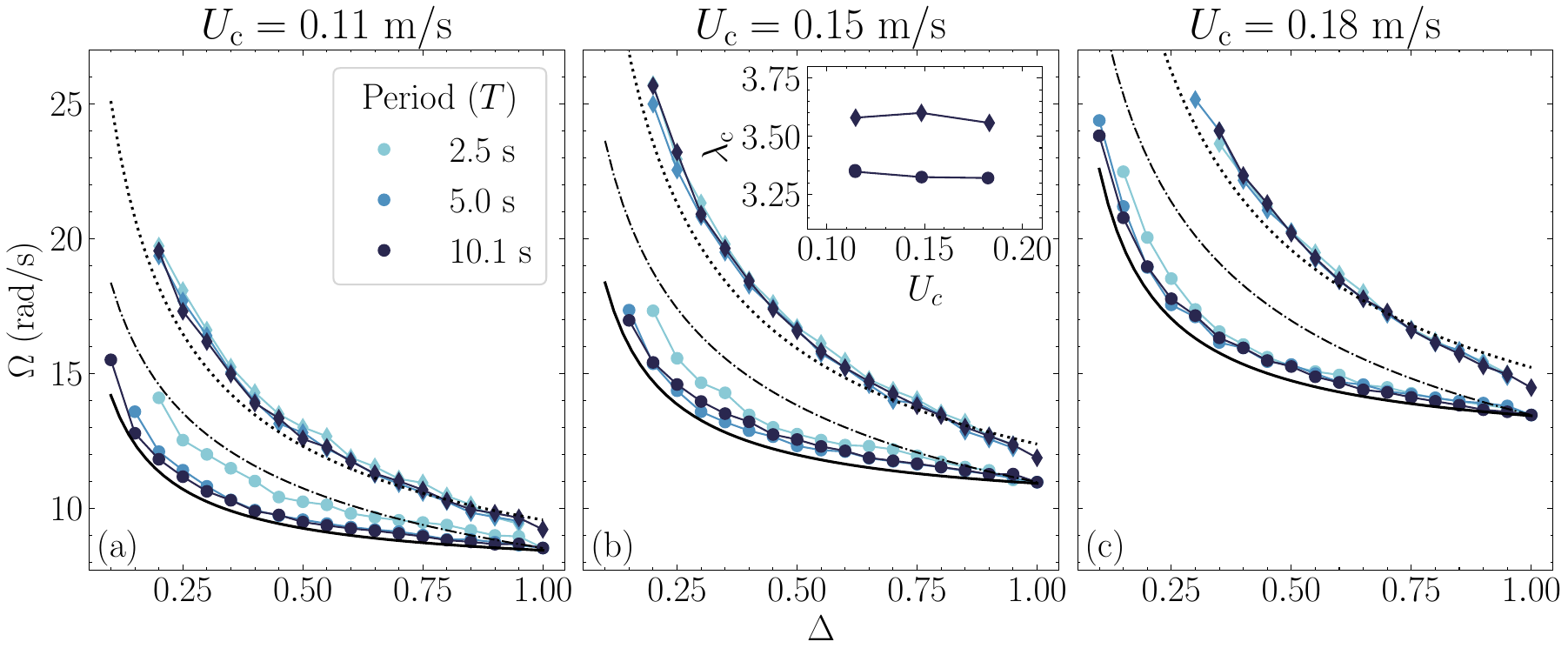}
    \caption{Angular velocity as a function of the duty cycle for three different periods (color). ($\blacklozenge$) Rigid propeller ($\bullet$) folding propeller. (full line) Angular velocity scaling from \eqref{eq:scaling_om_dc_flex} for the folding propeller.(dotted line) Identical scaling for the rigid propeller (dot-dashed line) Angular velocity scaling as $\Delta^{-1/3}$ for the folding propeller. A small shape discrepancy between the folding and rigid propellers causes the angular velocity required for continuous motion ($\Delta = 1$) to be slightly smaller for the folding propeller. This does not affect the power required significantly. The value for continuous motion is deduced from static force measurements (Fig. \ref{fig:perfo_rigvfold}). Inset : tip speed ratio of continuous motion $\lambda_c = R \Omega_c / U_\text{c}$ as a function of the mean velocity $U_\text{c}$ for both propellers.}
    \label{fig:oms_dc_3pan}
\end{figure*}

We compare the results of this reconfigurable propeller with those of the rigid propeller, whose shape corresponds to the fully open configuration by measuring the boat position, the propeller angular velocity, $\omega(t)$, and the hydrodynamic torque, $Q(t)$, as a function of time, $t$. This allows us to calculate the instantaneous power consumed in the rotation of the propeller, $P(t) = Q(t) \, \omega(t)$.
An experiment consists in finding pairs of duty cycle and angular velocity parameters such that, on average, the boat is stationary in the channel (the cruising speed of the boat corresponds to the fluid velocity, $U_\text{c}$, Fig. \ref{fig:setup}). We ran experiments for three different flow velocities (0.11, 0.15 and 0.18~m/s), three different periods (2.5, 5 and 10~s), and duty cycles ranging from 1 (continuous propulsion) to 0.1 (see Materials and Methods for details on the setup).

Aiming to find energy-efficient regimes, we choose the time periods, $T$, shorter than $t_{\text{max}}$, so that the boat does not lose too much of its momentum in the coast phase (or moves at the fluid velocity in the laboratory reference frame). $t_{\text{max}}$ corresponds to the balance between the inertia of the boat and its drag: $t_{\text{max}} \sim (m + m_a) / (1/2 \rho S_{\text{boat}} C_D U_\text{c})$, where $m$ is the mass of the boat, $m_a$ is the added mass due to the unsteady motion of the boat, $\rho$ is the fluid density, and $S_{\text{boat}}$ is the reference area of the boat (independent of the propeller configuration). For our experimental setup and parameters, $t_{\text{max}} \approx 10$ s.\\

For the three flow velocities, $U_\text{c}$, the three periods of boat motion, $T$, and for both types of propellers, rigid and folding, the angular velocity of the propeller, $\Omega$, required during the burst phase to keep the boat, on average, stationary in the flow increases as the duty cycle, $\Delta$, decreases (Fig. \ref{fig:oms_dc_3pan}).
The angular velocity is minimum for a duty cycle value of one, which corresponds to continuous rotation (no coast).
As the duty cycle decreases, the propeller produces thrust for a shorter time, so the required angular velocity increases accordingly.
Angular velocity shows little dependence on the period of motion compared to other parameters, {\it i.e.}, flow velocity, duty cycle, or type of propeller.
For a given duty cycle, the angular velocity increases with flow velocity; for continuous propulsion ($\Delta=1$), the tip speed ratio is constant ($R \Omega \propto U_\text{c}$, insert in Fig. \ref{fig:oms_dc_3pan}), which we show theoretically in the Appendix.
More importantly, the angular velocity of the folding propeller is always smaller than that of the rigid propeller. This difference increases as the duty cycle decreases, {\it i.e.}, as the duration of the coast phase increases (relative to the period duration). This is due to the reconfiguration of the folding propeller, which reduces the drag force during the coast phase and the averaged energy cost for motion.
More importantly, the angular velocity of the folding propeller is always smaller than that of the rigid propeller. This difference increases as the duty cycle decreases, {\it i.e.}, as the duration of the coast phase increases (relative to the period duration). This is due to the reconfiguration of the folding propeller, which reduces the drag force during the coast phase and the averaged energy cost for motion.
\begin{figure*}[htbp]
    \centering
    \includegraphics[width = 0.95 \linewidth]{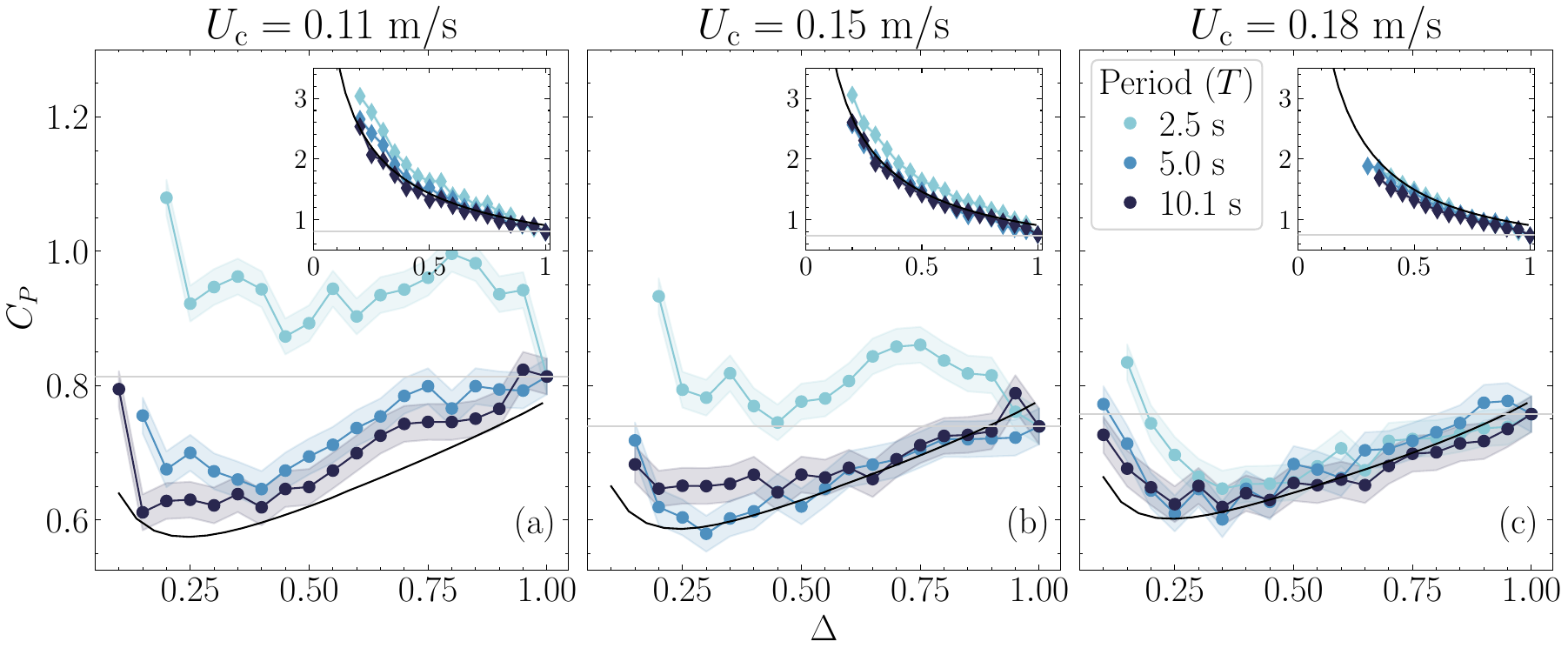}
    \caption{Power coefficient as a function of the duty cycle for the folding propeller for different periods: from light to dark $T = 2.5,\: 5$ and $10$ s. Panels (a), (b) and (c) are respectively associated with mean velocities $U_\text{c} = 0.11,\: 0.15$ and $0.18$ m/s. (dark line) Prediction from full model (described in Sec 3.\ref{sec:FullModel}) (light gray line) shows the value of the continuous regime ($\Delta = 1$). The shaded areas correspond to the uncertainty, estimated by repeating the experiments twice with $U_c = 0.11$~m/s and $T=5$~s.} Insets show the same graph, but for the rigid propeller (Note that vertical scale is different).
    \label{fig:cp_dc_3pan}
\end{figure*}
Assuming a propulsion force proportional to $\Omega^2$, if dissipation and propulsion efficiency are unaffected by burst and coast, and remain the same as for continuous motion, the energy expanded during a cycle is proportional to $\Omega^3 \Delta$ (constant work for a given cruising speed). The rotation velocity should therefore scale with $\Delta^{-1/3}$, which corresponds to the dotted dashed line in Figure \ref{fig:oms_dc_3pan}. The angular velocity is above this threshold for the rigid propeller, meaning that burst and coast is inefficient. However, the angular velocity for the folding propeller is below this threshold, so that intermittent motion may be more efficient than continuous activation.

We evaluate propulsion cost and efficiency using the dimensionless power coefficient $C_P$, which compares the average mechanical power supplied to the propeller, $\left< P \right>$, with the typical power dissipated by drag at a given cruising speed, $U_\text{c}$:
\begin{equation}\label{eq:cpdef}
    C_P = \frac{\left< P \right>}{\frac12 \rho S_\text{boat} U_\text{c}^3}.
\end{equation}
The mechanical power supplied by the propeller, $P(t)$, is the product of torque, $Q(t)$, and angular velocity $\omega(t)$, which we measure experimentally:
\begin{equation}\label{eq:pdef}
    \left< P \right> = \frac{1}{T} \int_0^T Q(t) \omega(t) \mathrm{d}t.
\end{equation}

With the rigid propeller, continuous propulsion is always more efficient than burst and coast; the power coefficient increases as the duty cycle decreases (Fig. \ref{fig:cp_dc_3pan}).
On the other hand, the intermittent propulsion is more efficient than the continuous propulsion  with the folding propeller over most of the parameter range. 
For a given cruising speed and period of motion, the power coefficient is minimum for an optimal duty cycle. For the three cruising speeds, the intermittent locomotion can reduce the energy cost by 15\%, reaching a maximum of 24\% for $U_\text{c} = 0.11 \, \rm{m/s}$, $T=10 \, \rm{s}$ at a duty cycle $\Delta \simeq 0.15$ (Fig. \ref{fig:cp_dc_3pan}a).

Short cycles seem less efficient. For the two lowest cruising speeds, $U_\text{c}=0.11 \, \rm{m/s}$ and $U_\text{c}=0.15 \, \rm{m/s}$, and for the shortest period, $T =2.5 \, {\rm s}$, the propulsion cost of intermittent rotation is larger than that of continuous rotation. We attribute this loss of efficiency to the cost of the rotational inertia of both the propeller and the fluid that the motor has to overcome when starting. 
This cost becomes a larger expense relative to the full cycle at smaller periods or cruising speeds. For the largest cruising speed, $U_\text{c}=0.18 \, \rm{m/s}$, the power coefficient shows little dependence on period for most values of duty cycle.

In order to explain and discuss the efficiency gain of intermittent motion and its optimum duty cycle, we now derive a model in the limit of small speed variations and negligible rotational inertia.

\section{Theoretical models explaining the gain in efficiency}\label{sec:Theo}
The hydrodynamic torque on the propeller, $Q$, is
\begin{equation}\label{eq:qdef}
    Q(U, \Omega) = \frac12 \rho \pi R^2 (R \Omega)^2 R C_Q(\lambda(t)),
\end{equation}
where $R$ is the radius of the rotor and $C_Q$ is the dimensionless torque coefficient, which {\it a priori} depends on the various dimensionless parameters, as well as the propeller design and geometry. In particular, $C_Q$ depends on the instantaneous tip speed ratio, $\lambda(t) = R \omega(t) / (U_\text{c} + U(t))$.  In our experiments (large Reynolds numbers), we find that for a given propeller, the torque depends only on the tip speed ratio and is constant in the limit of large tip speed ratios (Material and Methods, Fig. \ref{fig:perfo_rigvfold}). Equations \ref{eq:qdef}, \ref{eq:pdef}, and \ref{eq:cpdef} give:
\begin{equation}\label{eq:cp_expr_lbd}
    C_P = \frac{\pi R^2}{S_\text{boat}} \left( \frac{R \Omega}{U_\text{c}} \right)^3  \frac{1}{T} \int_0^{\Delta T} C_Q \left(\lambda(t) \right) \mathrm{d}t,
\end{equation}
where the rotation speed, $\Omega$, has to be determined as a function of the cruising speed and duty cycle.

\subsection{Simple scaling}\label{sec:SimpleModel}
We first derive a simple, analytical, scaling in the limit of no velocity variations ($\lambda(t) = {\rm cst}$ in the active phase and $\lambda(t) = 0$ in the passive phase) and constant torque coefficient ($C_Q=C_{ Q, 0}$).  In this limit, the power coefficient is simply:

\begin{equation}
    C_P = \frac{\pi R^5 C_{Q ,0}}{S_{\rm boat} U_\text{c}^3} \Delta \Omega^3,
\end{equation}

In steady state, the work of thrust exactly balances the energy dissipated by drag. Neglecting velocity variations ($25 \, \%$ at most in our experiments), and assuming constant drag, torque and efficiency coefficients, the balance gives 
\begin{multline}
    \frac{1}{2} \rho \pi R^2 (R \Omega)^2 C_{Q, 0} \, \eta \, \Delta = \\
\frac{1}{2} \rho S_{\rm{boat}} U_{\rm{c}}^2 \left[C_{\rm{D, \, B}} \Delta + C_{\rm{D, \, C}} (1-\Delta)\right],
\end{multline}
where $\eta$ is the propulsive efficiency which takes into account the conversion from torque to thrust (the typical efficiency of our propellers is 0.3). The thrust works over a distance $\Delta U_{\rm{c}} T$ and the drag force works over a distance $\Delta U_{\rm{c}} T$ with a coefficient $C_{D, \, \rm{B}}$ (burst, propeller in rotation) and over a distance $(1-\Delta) U_{\rm{c}} T$ with a coefficient $C_{D, \, \rm{C}}$ (coast, propeller at rest and folded when reconfigurable).
This simple balance gives the following scaling for the angular velocity, $\Omega$, with duty cycle, $\Delta$, and drag ratio, $C_{D, \text{C}}/C_{D, \text{B}}$ :
\begin{equation}\label{eq:scaling_om_dc_flex}
    \Omega = \Omega_c \sqrt{\frac{C_{D, \text{C}}}{C_{D, \text{B}}}\left( \frac{1}{\Delta} - 1 \right) + 1},
\end{equation}
where $\Omega_{\rm{c}}$ is the angular velocity of continuous motion ($\Delta=1$). As shown in Figure \ref{fig:oms_dc_3pan}, this scaling compares well with experiments, for which $C_{D, \text{C}}/C_{D, \text{B}} =0.2$ (full line) for the folding propeller and $C_{D, \text{C}}/C_{D, \text{B}} =0.65$ (dotted line) for the rigid propeller. The corresponding scaling for the power coefficient is
\begin{equation}
    C_{\rm{P}} = C_{\rm{P, c}} \Delta \left[ \frac{C_{D, \text{C}}}{C_{D, \text{B}}}\left( \frac{1}{\Delta} - 1 \right) + 1\right]^{3/2},
    \label{Cp_scaling}
\end{equation}
where $C_{\rm{P, c}}$ is the power coefficient of the continuous motion. Thanks to the drag reduction during the coast phase, the power coefficient first decreases when $\Delta$ decreases (longer coast) but diverges when $\Delta$ tends to zero because the rotation velocity concomitantly increases with $\Delta^{-1/2}$ (Eq. \ref{eq:scaling_om_dc_flex}). $C_{\rm{P}}$ as in equation \ref{Cp_scaling} is minimum for $\Delta = (C_{D, \text{C}}/C_{D, \text{B}})/[2(1-C_{D, \text{C}}/C_{D, \text{B}})]$, and is smaller than $C_{\rm{P, c}}$ (gain in efficiency) when the drag ratio, $C_{D, \text{C}}/C_{D, \text{B}}$, is smaller than $2/3$. For the drag ratio with the folding propeller ($C_{D, \text{C}}/C_{D, \text{B}} =0.2$), Eq. \ref{Cp_scaling} predicts an optimum at $\Delta =0.125$, which is comparable to the observed value in experiments ($\Delta \sim 0.15 \, - \, 0.3$, Fig. \ref{fig:cp_dc_3pan}) for a gain of $54 \%$, which is larger than the $15$ to $24 \%$ gain in experiments.

The correct order of magnitude of efficiency gain is obtained by taking into account the speed variations of the boat and, more importantly, the dependence of the torque coefficient on the tip speed ratio.

\subsection{Full model}\label{sec:FullModel}
To account for the velocity variations, we integrate the equations of motion, which are different for the coast and burst phases.\\
During the coast phase, only the drag force acts on the boat and slows it down. We model the drag force with two terms, the quasi-steady drag force and the added mass force:
\begin{equation}\label{eq:coast_drag}
    F_\text{drag} = \underbrace{\frac12 \rho S_\text{boat} U^2(t) C_{D, \text{C}}}_{\textrm{Quasi-steady}} + \underbrace{m_a \frac{\mathrm{d}U}{\mathrm{d}t}}_{\textrm{Added mass}}.
\end{equation}
We measured the quasi-steady drag coefficient of the boat, $C_{D, \text{C}}$, during the coast phase, i.e. when the propeller is not rotating, with both the folding and rigid propellers. We found them to be independent of the flow velocity when $U_{\rm c} > 0.1 {\rm m/s}$ (See appendix \ref{app:forces}). When not rotating, the folding propeller is fully closed, which reduces the drag by $80 \%$ compared to the rigid propeller.
The added mass is proportional to the volume of fluid displaced in front of the boat as it accelerates and increases the boat inertia. 
Thus, the equation of motion of the boat during the coast phase is:
\begin{equation}\label{eq:dim_coast}
    (m + m_a) \frac{\mathrm{d}U_1}{\mathrm{d}t} = - \frac12 \rho S_\text{boat} U_1^2(t) C_{D, \text{C}}.
\end{equation}

During the burst phase, the boat experiences both thrust and drag. These two forces are summed up in an axial force, $F_{\rm axial}$, where the thrust contribution is positive and scales as $\Omega^2$, while the drag contribution is negative and scales as $U_\text{c}^2$:
\begin{equation}
    F_{\rm axial} = \frac12 \rho \pi R^2 (R \Omega)^2 C_a(\lambda),
\end{equation}
where $C_a$ is a function of $\lambda$: $C_a(\lambda) = a + c / \lambda^2$. The $a$ coefficient represents the thrust of the propeller and is positive, while the $c$ coefficient represents the quasi-steady drag of the boat and propeller and is negative ($C_{D, B} = -\pi R^2 / S_{\text{boat}} c$). We have determined the values of these two coefficients experimentally in static conditions and make the hypothesis that they remain unchanged during unsteady motion (See Supplementary Information). 

With the added mass, the equation of motion in the burst phase is: 
\begin{equation}
    (m + m_a) \frac{\mathrm{d}U_2}{\mathrm{d}t} = \frac12 \rho \pi R^2 (R \Omega)^2 \left( a + \frac{c}{\lambda^2} \right).
\end{equation}

We make our system dimensionless by rescaling the time by the period $T$ and the velocities by $R \Omega$. We denote the dimensionless velocities by a lower case u and the dimensionless time derivative by a dot. Our set of equations becomes :
\begin{equation}\label{eq:dimles_coast}
    \dot{u_1} = - M \frac{S_\text{boat}}{\pi R^2}C_{D, \text{C}} u_1^2,
\end{equation}
during the coast phase and :
\begin{equation}\label{eq:dimles_burst}
    \dot{u_2} = M \left( a + c u_2^2 \right),
\end{equation}
during the burst phase, where $M$ is a dimensionless coefficient that compares the mass of fluid moved by the rotating propeller to that of the system in translation:
\begin{equation}
    M = T \Omega \frac{\rho \pi R^3 }{2(m + m_a)}.
\end{equation}
Looking for steady-state, periodic solutions, the initial conditions are
\begin{equation}\label{eq:dimles_bc}
    \quad u_1(1) = u_2(0), \ u_2(\Delta) = u_1(\Delta).
\end{equation}

We solve this system numerically to find the rotation velocity, $\Omega$, or the value of $M$, so that the boat moves at the desired cruising speed $U_{\rm c}$ for a given duty cycle, $\Delta$. Knowing the rotation velocity, $\Omega$, and the time-dependent velocity of motion $U(t)$, we calculate the time-dependent tip speed ratio, $\lambda(t)$, to integrate the power coefficient (Eq \ref{eq:cp_expr_lbd}), using the torque coefficient $C_{Q} (\lambda(t))$ that we determined experimentally in static conditions. Just as we have modelled axial force as the sum of the thrust and drag contributions, we model torque as the sum of a component resisting rotation (propulsion, proportional to $\Omega^2$) and a component in the same direction as the rotation (turbine, proportional to $U_\text{c}^2$), $C_Q = a_Q + c_Q / \lambda^2$ (see Materials and Methods, Fig. \ref{fig:perfo_rigvfold}). 

This model agrees well with experimental results for long period, as shown in Fig. \ref{fig:cp_dc_3pan} ($C_{\rm P} (\Delta)$) for $T=10 \, s$. However, this model predicts that the gain increases ({\it i.e.} $C_P$ decreases), as the period (or $M$) decreases, contrary to what is observed in experiments. We believe this discrepancy is due to transient effects such as rotational inertia, which are not included in our model. Unsteady wave drag could also play a significant role. Depending on the period, wave drag has been shown to either increase or decrease the mean drag \citep{dode_wave_2022}.

Since the quasi-steady limit (long period, low cruise speed) is the most efficient, we discuss the parameters that give peak efficiency and compare propeller propulsion and fish swimming in the light of this model.

\section{Discussion}
To discuss the efficiency gain of intermittent propulsion, we vary the main parameters of the system, the flow velocity and drag ratio, while keeping the period ($T = 10 \, \rm{s}$) and geometric parameters constant, and find the best theoretical relative efficiency gain:
\begin{equation}\label{eq:dcpdef}
    \tau = - \min_\Delta \frac{C_P(\Delta) - C_P(1)}{C_P(1)},
\end{equation}
and the corresponding duty cycle:
\begin{equation}\label{eq:dmaxdef}
    \Delta_\text{max} = \underset{\Delta}{\text{argmin}} \frac{C_P(\Delta) - C_P(1)}{C_P(1)}.
\end{equation}

\begin{figure}[htbp]
    \centering
    \includegraphics[width = \linewidth]{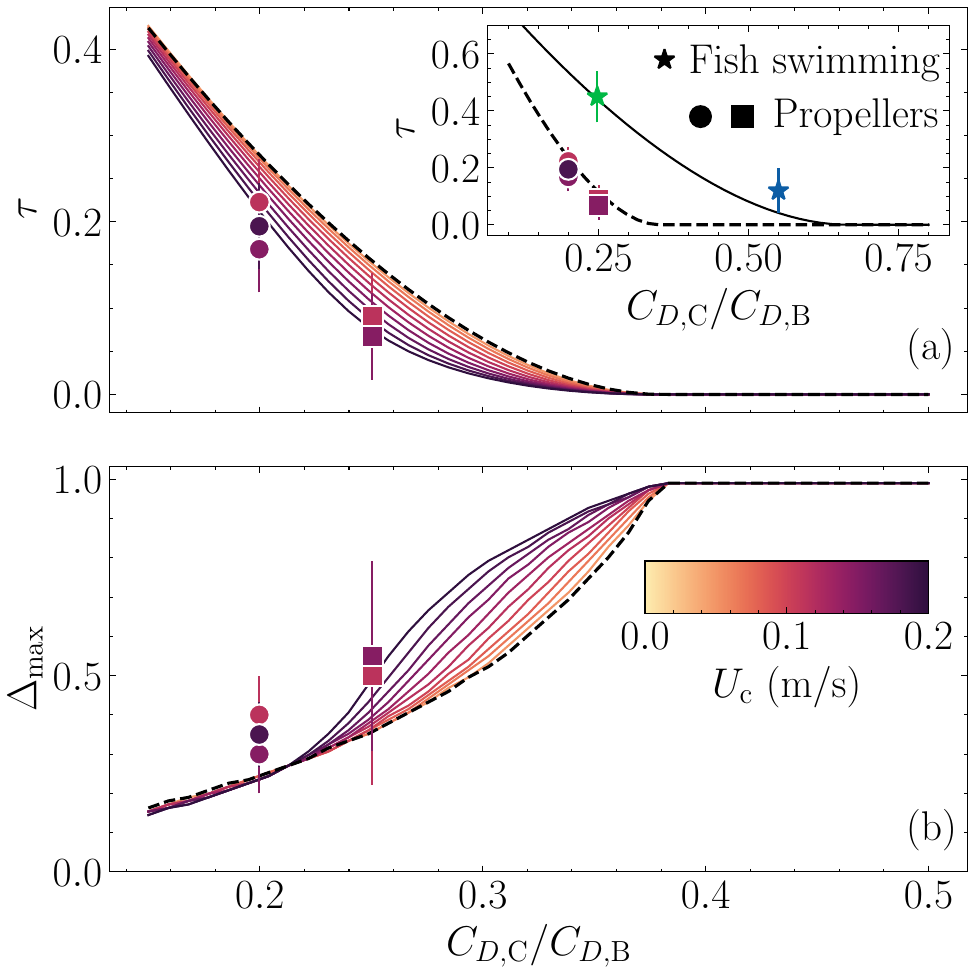}
    \caption{(a) Maximal relative mean power gains compared to continuous motion as a function of the drag ratio between the coast and burst phases. Error bars correspond to uncertainty in the data, which are estimated by repeating experiments. Inset: Comparison with fish swimming literature. (Blue) Pitching airfoil experiments from Floryan et al., 2017 \cite{floryan_forces_2017} (Green) Experiments with carps from Wu et al., 2007 \cite{wu_kinematics_2007} (b) Duty cycle for which the power reduction is maximum. Error bars correspond to the range where the power coefficient is less than 5\% away from the optimum. Data from experiments with water height ($\bullet$) 190 mm, ($\blacksquare$) 200 mm (See Fig. \ref{fig:cp_dc_3pan}). (dashed line) Solution from Eq. \ref{eq:cp_analytic}. (full line) Solution with a constant torque coefficient (Eq. \ref{eq:cp_analytic_cq_cst}).} 
    \label{fig:simu_optim}
\end{figure}

As shown in Figure \ref{fig:simu_optim}, for all flow velocities (from 0 to 0.2 m/s), intermittent propulsion is more efficient than continuous propulsion when the drag ratio, $C_{D, \text{C}} / C_{D, \text{B}}$, is less than 0.35. Then the gain increases and the optimum duty cycle decreases as the drag ratio decreases. These theoretical results are confirmed by our experiments for two different drag ratios.
In the experiments, we vary the drag ratio by changing the relative importance of propeller and hull drag. The hull drag was increased by increasing the water height and the immersion of the hull. For a drag ratio of 0.2, the efficiency gain is between 15\% and 24\% for an optimum duty cycle of $0.35 \pm 0.5$, depending on the flow velocity. When the drag ratio increases to 0.25, the optimum efficiency gain is only 7.5 \% for a duty cycle of 0.55.

 In the model, the efficiency gain increases as the flow velocity decreases.  This is due to boat speed variations, which increase with mean coast velocity (flow velocity) and are detrimental to efficiency. 
This trend is not clearly observed in our experiments, probably because other sources of dissipation, such as rotational inertia, mask the effect of speed variations.
As discussed in Section \ref{sec:SimpleModel}, in the limit of no speed variation ({\it i.e.} large boat inertia or $M \to 0$), the relative power gain reduces to:
\begin{equation}\label{eq:cp_analytic}
   \tau=1 - \left[ \frac{C_{\rm D,C}}{C_{\rm D,B}} \left( \frac{1}{\Delta_{\rm max}} -1 \right) +1 \right]^{3/2} \frac{C_{Q} (\lambda)}{C_{Q, 0}} ,
\end{equation}
where the torque coefficient $C_{Q}(\lambda)$ is a function of the tip speed ratio ($C_Q = a_Q + c_Q / \lambda^2$, Fig. \ref{fig:perfo_rigvfold}) and therefore of the duty cycle, $\Delta$ (the tip speed ratio, $\lambda= R \Omega / U_{\rm c}$, is a function of $\Delta$ as defined by \eqref{eq:scaling_om_dc_flex}). $C_{Q, 0}$ is the value of the torque coefficient in the continuous regime. This limit also corresponds to the limit of small flow velocities (dashed line, Fig. \ref{fig:simu_optim}), where the energy saving is slightly larger than experimental values.

The dependence of the torque coefficient on the tip ratio has a major influence on the relative efficiency gain. 
Since the torque coefficient of the propeller increases with the tip speed ratio (Fig. \ref{fig:perfo_rigvfold}), which is inevitably larger in the intermittent mode than in the continuous mode, the relative efficiency gain is less than for a constant torque coefficient, for which:
\begin{equation}\label{eq:cp_analytic_cq_cst}
   \tau=1- \frac{3}{2}  \frac{C_{D, \text{C}}}{C_{D, \text{B}}} \sqrt{3 \left(1 -  \frac{C_{D, \text{C}}}{C_{D, \text{B}}} \right)} \ {\rm when} \ \frac{C_{D, \text{C}}}{C_{D, \text{B}}} < \frac{2}{3},
\end{equation}
as shown in the inset in Fig. \ref{fig:simu_optim}.   
In particular, intermittent propulsion is favorable for a drag ratio less than $0.35$ compared to $2/3$ when the torque coefficient is independent of the tip speed ratio. This more favorable regime seems to be found in fish, as shown in the inset of Fig. \ref{fig:simu_optim}, where we have plotted data from previous studies with live carp \citep{wu_kinematics_2007} and pitching airfoil propulsion \citep{floryan_scaling_2017} ($Re \sim \mathcal{O}(10^3 - 10^4)$ as in our study).
Indeed, experimental data from Floryan {\it et al.} \cite{floryan_scaling_2017} show that the dimensionless cost of locomotion (equivalent to $C_Q$), when using a pitching airfoil, is independent of the actuation parameter (the Strouhal number, $St = f A / U_\text{c}$ where $f$ and $A$ are the tail-beat frequency and amplitude, respectively, analogous to the tip speed ratio).
This particular dependence of the dimensionless cost of locomotion on the actuation parameter makes intermittent pitching particularly advantageous in fish propulsion. For a drag ratio of 0.25, the relative efficiency gain for carp is of 45\% \cite{wu_kinematics_2007}. With a drag ratio of 0.55, intermittent propulsion is still beneficial ($\tau = 0.1$) for pitching airfoils \cite{floryan_forces_2017} (Fig. \ref{fig:simu_optim}).

\section{Conclusion}\label{sec:conclusion}

We have experimentally investigated the burst and coast dynamics in the case of a reconfigurable propeller, bio-inspired by fish swimming. Our work demonstrates that energy savings through intermittent motion can be achieved passively in artificial devices such as a propeller-driven boat. The conclusions of this work are applicable to any system driven through a fluid and subject to a typical high Reynolds number drag force that is quadratic in the speed of motion, such as fish or human swimming, bird flight, etc.
Intermittent motion is advantageous when the drag ratio between the coast and burst phases is less than a threshold value, which is critically determined by the dependence of thrust cost on the actuation number, which compares the characteristic driving velocity to the moving speed, e.g. the tip speed number for propeller propulsion or the Strouhal number for fish swimming. For our propeller driven boat, for which cost of thrust increases with the tip speed ratio, intermittent motion is advantageous for drag ratios less than 0.35. The onset increases to 2/3 when the thrust cost is independent of the actuation number, which is the case for pitching propulsion as for fish. A decreasing thrust cost with the actuation number would make burst and cost motion even more advantageous.

\newpage
\appendix

\section{Materials and Methods}

\subsection*{Intermittent propulsion setup}

The boat is driven by a DC motor whose angular velocity is controlled to follow a prescribed time sequence. The torque is measured using a mechanical sensor mounted between the motor shaft and the propeller. Before each experiment, we measured the torque as a function of the angular velocity of the system without propeller, which we subtracted from the torque measured during the experiment (with the propeller) to only retain the hydrodynamic part.
The boat is free to move along a linear air bearing rail (of negligible friction) in a 1.3 meter-long water channel of rectangular section $0.19 \, \times \, 0.2 \, {\rm m}$. 
The position of the boat is measured with an ultrasonic sensor.
The water flow in the channel is generated by a pump of variable speed. The mean flow velocity, ranging from $0.05$ to $0.25$ m/s, is derived from measurements of water height and flux. A settling chamber, followed by a contraction and honeycomb flow straighteners dampen turbulent fluctuations at the channel entry. The Reynolds number based on the size of the hull is about 15~000.

During an experiment, we fixed the water speed and set the time sequence of the electric motor to a series of (quasi-) square waves with a given time period, duty cycle and speed. The square waves were convolved with a gaussian to smooth the motor accelerations (Fig. \ref{fig:setup} ).  
After a transient regime, the boat travelled either upstream or downstream at a constant speed. The speed of the motor was then trimmed so that the position of the boat did not change by more than $2$ cm for at least $10$ cm in a row. The same experiments were repeated several times resulting in a 1\% variation in angular velocity and a 3.5\% variation in the power coefficient.

\begin{figure}[htbp]
    \centering
    \includegraphics[width = 0.95 \linewidth]{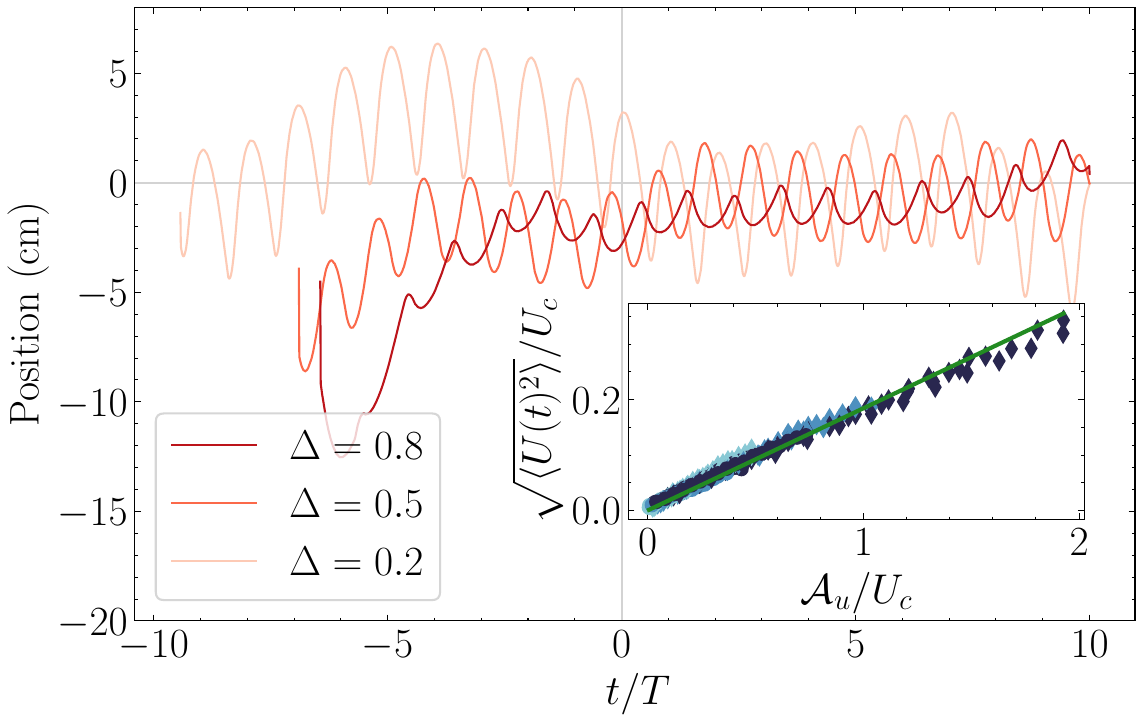}
    \caption{Three examples of boat trajectories in the laboratory reference frame. Colors represent the duty cycles (darker is longer). Here the propeller is rigid, $U_\text{c} = 0.15$ m/s, and $T = 5$ s. From $t = 0$ the position of the boat is (quasi-) stationary on average. (Inset) Root-mean-square fluctuations of the velocity as a function of $\mathcal{A}_u$. The green line is a fit with slope 0.18.}
    \label{fig:example_tempo}
\end{figure}

The speed of the boat in the fluid reference frame is $U_\text{c} + U(t)$, where $U_\text{c}$ is the fluid velocity in the laboratory frame and $U(t)$ is the speed of the boat in the laboratory frame, which for the reported values is (quasi-) zero on average (Fig. \ref{fig:example_tempo}). This transformation of frames of reference is Galilean.

Given the intermittent nature of the motion, the velocity varies over time (Fig. \ref{fig:example_tempo}). The amplitude of the velocity fluctuation $\mathcal{A}_u$, can be estimated using Eq. \eqref{eq:dim_coast}:
\begin{equation}\label{eq:vfluc}
    \mathcal{A}_u = (1 - \Delta) T \frac{1/2 \rho S_\textrm{boat}C_\textrm{D,C}U_c^2}{m + m_a}.
\end{equation}
Eq. \eqref{eq:vfluc} fits well with the experimental data (inset of Fig. \ref{fig:example_tempo}), collapsing data for different propellers, mean velocities, duty cycles, and periods.

\subsection*{Forces measurements}\label{app:forces}
In order to derive the drag and thrust coefficients of the boat, we measured the horizontal force, $D$, acting on the boat when subjected to a water flow with or without rotating the propeller. For these specific experiments, the boat is fixed and connected to a horizontal force sensor. 

\begin{figure}[htbp]
    \centering
    \includegraphics[width = 0.95\linewidth]{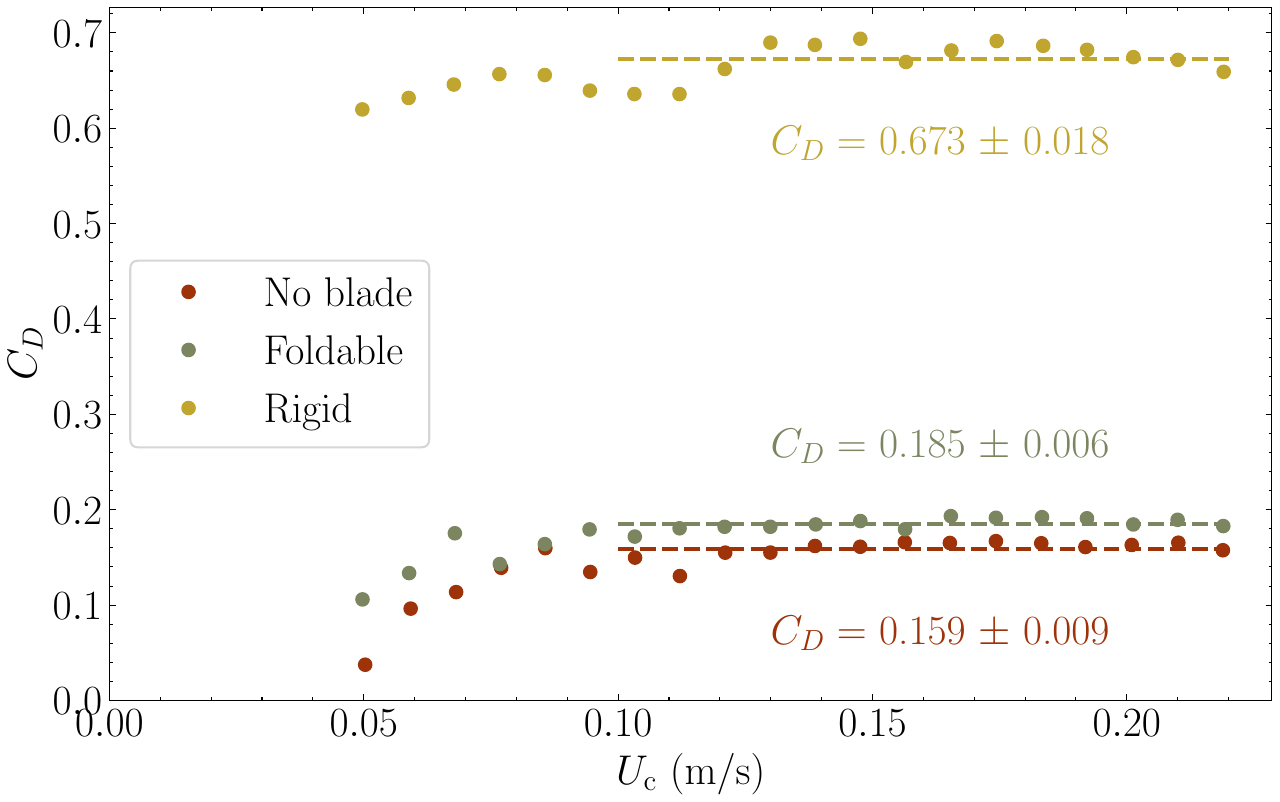}
    \caption{Drag coefficient as a function of the flow speed. (\textcolor{blue}{$\bullet$}) boat hull without propeller, (\textcolor{green}{$\bullet$}) with the reconfigurable propeller completely folded, and (\textcolor{orange}{$\bullet$}) with the rigid propeller. In intermittent propulsion experiments, the flow velocity is greater than or equal to $0.11$ m/s. In the calculations we use the reported plateau values for the drag coefficients during the coast phase, $C_{\rm D, C}$.}
    \label{fig:drag}
\end{figure}

We measured the drag coefficient, $C_{\rm D}$, of the boat as a function of the flow velocity, $U_{\rm c}$, with the rigid propeller, the reconfigurable propeller when folded and without propeller (Fig. \ref{fig:drag}). To measure $C_{D}$, propellers do not rotate. The drag coefficient is defined as $C_D = D / (1/2 \rho S_{\rm boat} U_\text{c}^2)$ where $S_{\rm boat}$ is the boat surface of reference, $S = 9.6\:10^{-3}$ m$^2$.

\begin{figure}[tbp]
    \centering
    \includegraphics[width = 0.95\linewidth]{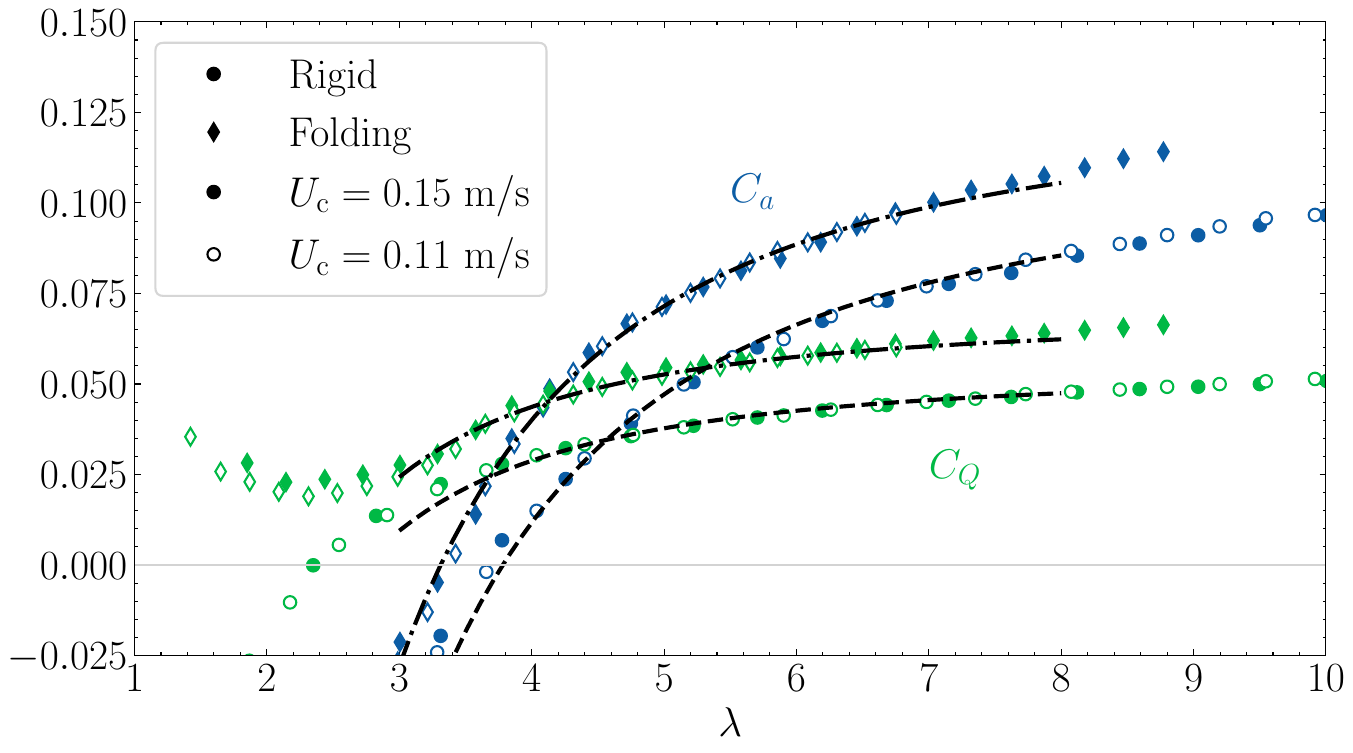}
    \caption{Efficiency of propellers. Axial force coefficient (blue), $C_{\rm a}$, and torque coefficient (green), $C_{Q}$, of the ($\bullet$) rigid and ($\blacklozenge$) flexible propellers as a function of the tip speed ratio, $\lambda$, for two flow velocities. 
    There is a visible offset
between data of the rigid and the folding propeller. We believe that the hinges of the reconfigurable blades introduce a slight tilt, which increases the angle of attack of the blades.
The curves show the best fits (see text).}
    \label{fig:perfo_rigvfold}
\end{figure}

We measured the axial force, $F_{\rm axial}$, (the sum of thrust and drag) and the torque, $Q$, as a function of the rotation velocity of propellers for two different flow velocity, $U_{\rm c}=0.11 \, {\rm m/s}$ and  $U_{\rm c}=0.15 \, {\rm m/s}$, and computed the axial force coefficient, $C_{\rm a}$, and the torque coefficient, $C_{Q}$:
\begin{equation}
    C_a = \dfrac{F_\text{axial}}{\frac12 \rho \pi R^2 \left( R \Omega \right)^2}, \quad C_Q = \dfrac{Q}{\frac12 \rho \pi R^2 \left( R \Omega \right)^2 R}.
\end{equation}
For both the rigid and the folding propeller, $C_{\rm a}$ and $C_{Q}$ are only functions of the tip speed ratio $\lambda$ (Fig. \ref{fig:perfo_rigvfold}).
Their dependencies on tip speed ratio are typical of marine propellers: they first increase $\lambda$ and tend to a constant value at high $\lambda$.  
Experimental data are fitted with equations $C_a = a + c / \lambda^2$ and $C_Q = a_Q + c_Q / \lambda^2$ in the range $3.5 < \lambda < 8$ (Fig. \ref{fig:perfo_rigvfold} and Table \ref{tab:fits}).
$a$ and $c$ coefficients represent thrust and drag, respectively. 
$a_Q$ and $c_Q$ coefficients represent propulsion and turbine behavior of propellers, respectively.

\begin{table}
    \centering
    \begin{tabular}{| c c c c c |}
        \hline
       Propeller& $a$ & $c$ & $a_Q$ & $c_Q$\\
        \hline
       Rigid & +0.1111 & -1.6065 & +0.0545 & -0.4196 \\
       Folding & +0.1268 & -1.3777 & +0.0689 & -0.4082 \\
       Prop. 4381 \cite{boswell_design_1971} & + 0.1452 & -1.6259 & +0.0492 & -0.4066 \\
        \hline
    \end{tabular}
    \caption{Fitting parameters of the axial force and torque coefficients, for both propellers used in experiments (Fig. \ref{fig:perfo_rigvfold}). Third line corresponds to the parameter values of an highly efficient marine propeller design \cite{boswell_design_1971}.}
    \label{tab:fits}
\end{table}

\subsection*{Equations of motion}
Integrating equations of motion in both phases, \eqref{eq:dimles_coast}) (coast) and (\eqref{eq:dimles_burst} (burst), gives: 
\begin{equation}
    \begin{split}
        u_1(t) &= \left( A + M\frac{S_\text{boat}}{\pi R^2} C_{D, \text{C}} (t - \Delta) \right)^{-1}, \\
         u_2(t) &= \sqrt{\frac{-a}{c}} \tanh\left( \sqrt{-ac} \left(B + M t \right)\right),
    \end{split}
\end{equation}
where $A$ and $B$ are integration constants, which are found numerically (root finding algorithm) to satisfy boundary conditions: the final velocity of one phase is the starting velocity of the other.
For both types of propellers, we ran the simulations for the same parameters as in the experiments and for a period of 10 s (Fig. \ref{fig:cp_dc_3pan}).
As in the experiment, we looked for pairs $(\Delta, \Omega)$ such that the mean speed of the boat was equal to the prescribed value.
Finally, we have verified that the main results presented in this manuscript are general and not strongly dependent on the propeller design.
For a highly efficient marine propeller (parameters in Table \ref{tab:fits} \cite{boswell_design_1971}), the model (\eqref{eq:cp_analytic}) shows that intermittent motion is advantageous for drag ratios less than 0.3 (0.35 with the parameters of the reconfigurable propeller used in this study). For a drag ratio of 0.2, optimum power consumption is reduced by about 20\% compared with continuous motion.


\begin{thebibliography}{25}%
\makeatletter
\providecommand \@ifxundefined [1]{%
 \@ifx{#1\undefined}
}%
\providecommand \@ifnum [1]{%
 \ifnum #1\expandafter \@firstoftwo
 \else \expandafter \@secondoftwo
 \fi
}%
\providecommand \@ifx [1]{%
 \ifx #1\expandafter \@firstoftwo
 \else \expandafter \@secondoftwo
 \fi
}%
\providecommand \natexlab [1]{#1}%
\providecommand \enquote  [1]{``#1''}%
\providecommand \bibnamefont  [1]{#1}%
\providecommand \bibfnamefont [1]{#1}%
\providecommand \citenamefont [1]{#1}%
\providecommand \href@noop [0]{\@secondoftwo}%
\providecommand \href [0]{\begingroup \@sanitize@url \@href}%
\providecommand \@href[1]{\@@startlink{#1}\@@href}%
\providecommand \@@href[1]{\endgroup#1\@@endlink}%
\providecommand \@sanitize@url [0]{\catcode `\\12\catcode `\$12\catcode `\&12\catcode `\#12\catcode `\^12\catcode `\_12\catcode `\%12\relax}%
\providecommand \@@startlink[1]{}%
\providecommand \@@endlink[0]{}%
\providecommand \url  [0]{\begingroup\@sanitize@url \@url }%
\providecommand \@url [1]{\endgroup\@href {#1}{\urlprefix }}%
\providecommand \urlprefix  [0]{URL }%
\providecommand \Eprint [0]{\href }%
\providecommand \doibase [0]{https://doi.org/}%
\providecommand \selectlanguage [0]{\@gobble}%
\providecommand \bibinfo  [0]{\@secondoftwo}%
\providecommand \bibfield  [0]{\@secondoftwo}%
\providecommand \translation [1]{[#1]}%
\providecommand \BibitemOpen [0]{}%
\providecommand \bibitemStop [0]{}%
\providecommand \bibitemNoStop [0]{.\EOS\space}%
\providecommand \EOS [0]{\spacefactor3000\relax}%
\providecommand \BibitemShut  [1]{\csname bibitem#1\endcsname}%
\let\auto@bib@innerbib\@empty
\bibitem [{\citenamefont {Kramer}\ and\ \citenamefont {McLaughlin}(2001)}]{kramer_behavioral_2001}%
  \BibitemOpen
  \bibfield  {author} {\bibinfo {author} {\bibfnamefont {D.~L.}\ \bibnamefont {Kramer}}\ and\ \bibinfo {author} {\bibfnamefont {R.~L.}\ \bibnamefont {McLaughlin}},\ }\bibfield  {title} {\bibinfo {title} {The {Behavioral} {Ecology} of {Intermittent} {Locomotion}},\ }\href {https://doi.org/10.1093/icb/41.2.137} {\bibfield  {journal} {\bibinfo  {journal} {American Zoologist}\ }\textbf {\bibinfo {volume} {41}},\ \bibinfo {pages} {137} (\bibinfo {year} {2001})}\BibitemShut {NoStop}%
\bibitem [{\citenamefont {Tobalske}(2010)}]{tobalske_hovering_2010}%
  \BibitemOpen
  \bibfield  {author} {\bibinfo {author} {\bibfnamefont {B.~W.}\ \bibnamefont {Tobalske}},\ }\bibfield  {title} {\bibinfo {title} {Hovering and intermittent flight in birds},\ }\href {https://doi.org/10.1088/1748-3182/5/4/045004} {\bibfield  {journal} {\bibinfo  {journal} {Bioinspiration \& Biomimetics}\ }\textbf {\bibinfo {volume} {5}},\ \bibinfo {pages} {045004} (\bibinfo {year} {2010})}\BibitemShut {NoStop}%
\bibitem [{\citenamefont {Rayner}\ \emph {et~al.}(2001)\citenamefont {Rayner}, \citenamefont {Viscardi}, \citenamefont {Ward},\ and\ \citenamefont {Speakman}}]{rayner_aerodynamics_2001}%
  \BibitemOpen
  \bibfield  {author} {\bibinfo {author} {\bibfnamefont {J.~M.~V.}\ \bibnamefont {Rayner}}, \bibinfo {author} {\bibfnamefont {P.~W.}\ \bibnamefont {Viscardi}}, \bibinfo {author} {\bibfnamefont {S.}~\bibnamefont {Ward}},\ and\ \bibinfo {author} {\bibfnamefont {J.~R.}\ \bibnamefont {Speakman}},\ }\bibfield  {title} {\bibinfo {title} {Aerodynamics and {Energetics} of {Intermittent} {Flight} in {Birds}},\ }\href@noop {} {\bibfield  {journal} {\bibinfo  {journal} {American Zoologist}\ } (\bibinfo {year} {2001})}\BibitemShut {NoStop}%
\bibitem [{\citenamefont {Wu}\ \emph {et~al.}(2007)\citenamefont {Wu}, \citenamefont {Yang},\ and\ \citenamefont {Zeng}}]{wu_kinematics_2007}%
  \BibitemOpen
  \bibfield  {author} {\bibinfo {author} {\bibfnamefont {G.}~\bibnamefont {Wu}}, \bibinfo {author} {\bibfnamefont {Y.}~\bibnamefont {Yang}},\ and\ \bibinfo {author} {\bibfnamefont {L.}~\bibnamefont {Zeng}},\ }\bibfield  {title} {\bibinfo {title} {Kinematics, hydrodynamics and energetic advantages of burst-and-coast swimming of koi carps ({Cyprinus} carpio koi)},\ }\href@noop {} {\bibfield  {journal} {\bibinfo  {journal} {The Journal Of Experimental Biology}\ }\textbf {\bibinfo {volume} {210}},\ \bibinfo {pages} {2181} (\bibinfo {year} {2007})}\BibitemShut {NoStop}%
\bibitem [{\citenamefont {Ribak}\ \emph {et~al.}(2005)\citenamefont {Ribak}, \citenamefont {Weihs},\ and\ \citenamefont {Arad}}]{ribak_submerged_2005}%
  \BibitemOpen
  \bibfield  {author} {\bibinfo {author} {\bibfnamefont {G.}~\bibnamefont {Ribak}}, \bibinfo {author} {\bibfnamefont {D.}~\bibnamefont {Weihs}},\ and\ \bibinfo {author} {\bibfnamefont {Z.}~\bibnamefont {Arad}},\ }\bibfield  {title} {\bibinfo {title} {Submerged swimming of the great cormorant \textit{{Phalacrocorax} carbo sinensis} is a variant of the burst-and-glide gait},\ }\href {https://doi.org/10.1242/jeb.01856} {\bibfield  {journal} {\bibinfo  {journal} {Journal of Experimental Biology}\ }\textbf {\bibinfo {volume} {208}},\ \bibinfo {pages} {3835} (\bibinfo {year} {2005})}\BibitemShut {NoStop}%
\bibitem [{\citenamefont {Xia}\ \emph {et~al.}(2018)\citenamefont {Xia}, \citenamefont {Chen}, \citenamefont {Liu},\ and\ \citenamefont {Luo}}]{xia_energy-saving_2018}%
  \BibitemOpen
  \bibfield  {author} {\bibinfo {author} {\bibfnamefont {D.}~\bibnamefont {Xia}}, \bibinfo {author} {\bibfnamefont {W.-s.}\ \bibnamefont {Chen}}, \bibinfo {author} {\bibfnamefont {J.-k.}\ \bibnamefont {Liu}},\ and\ \bibinfo {author} {\bibfnamefont {X.}~\bibnamefont {Luo}},\ }\bibfield  {title} {\bibinfo {title} {The energy-saving advantages of burst-and-glide mode for thunniform swimming},\ }\href {https://doi.org/10.1007/s42241-018-0120-8} {\bibfield  {journal} {\bibinfo  {journal} {Journal of Hydrodynamics}\ }\textbf {\bibinfo {volume} {30}},\ \bibinfo {pages} {1072} (\bibinfo {year} {2018})}\BibitemShut {NoStop}%
\bibitem [{\citenamefont {Coughlin}\ \emph {et~al.}(2022)\citenamefont {Coughlin}, \citenamefont {Chrostek},\ and\ \citenamefont {Ellerby}}]{coughlin_intermittent_2022}%
  \BibitemOpen
  \bibfield  {author} {\bibinfo {author} {\bibfnamefont {D.~J.}\ \bibnamefont {Coughlin}}, \bibinfo {author} {\bibfnamefont {J.~D.}\ \bibnamefont {Chrostek}},\ and\ \bibinfo {author} {\bibfnamefont {D.~J.}\ \bibnamefont {Ellerby}},\ }\bibfield  {title} {\bibinfo {title} {Intermittent propulsion in largemouth bass, {Micropterus} salmoides, increases power production at low swimming speeds},\ }\href {https://doi.org/10.1098/rsbl.2021.0658} {\bibfield  {journal} {\bibinfo  {journal} {Biology Letters}\ }\textbf {\bibinfo {volume} {18}},\ \bibinfo {pages} {20210658} (\bibinfo {year} {2022})}\BibitemShut {NoStop}%
\bibitem [{\citenamefont {Videler}(1981)}]{videler_swimming_1981}%
  \BibitemOpen
  \bibfield  {author} {\bibinfo {author} {\bibfnamefont {J.}~\bibnamefont {Videler}},\ }\bibfield  {title} {\bibinfo {title} {Swimming {Movements}, {Body} {Structure} and {Propulsion} in {Cod} {Gadus} morhua},\ }\href@noop {} {\bibfield  {journal} {\bibinfo  {journal} {Symposia of the Zoological Society of London}\ }\textbf {\bibinfo {volume} {48}},\ \bibinfo {pages} {1} (\bibinfo {year} {1981})}\BibitemShut {NoStop}%
\bibitem [{\citenamefont {Naemi}\ \emph {et~al.}(2010)\citenamefont {Naemi}, \citenamefont {Easson},\ and\ \citenamefont {Sanders}}]{naemi_hydrodynamic_2010}%
  \BibitemOpen
  \bibfield  {author} {\bibinfo {author} {\bibfnamefont {R.}~\bibnamefont {Naemi}}, \bibinfo {author} {\bibfnamefont {W.~J.}\ \bibnamefont {Easson}},\ and\ \bibinfo {author} {\bibfnamefont {R.~H.}\ \bibnamefont {Sanders}},\ }\bibfield  {title} {\bibinfo {title} {Hydrodynamic glide efficiency in swimming},\ }\href {https://doi.org/10.1016/j.jsams.2009.04.009} {\bibfield  {journal} {\bibinfo  {journal} {Journal of Science and Medicine in Sport}\ }\textbf {\bibinfo {volume} {13}},\ \bibinfo {pages} {444} (\bibinfo {year} {2010})}\BibitemShut {NoStop}%
\bibitem [{\citenamefont {Barbosa}\ \emph {et~al.}(2010)\citenamefont {Barbosa}, \citenamefont {Bragada}, \citenamefont {Reis}, \citenamefont {Marinho}, \citenamefont {Carvalho},\ and\ \citenamefont {Silva}}]{barbosa_energetics_2010}%
  \BibitemOpen
  \bibfield  {author} {\bibinfo {author} {\bibfnamefont {T.~M.}\ \bibnamefont {Barbosa}}, \bibinfo {author} {\bibfnamefont {J.~A.}\ \bibnamefont {Bragada}}, \bibinfo {author} {\bibfnamefont {V.~M.}\ \bibnamefont {Reis}}, \bibinfo {author} {\bibfnamefont {D.~A.}\ \bibnamefont {Marinho}}, \bibinfo {author} {\bibfnamefont {C.}~\bibnamefont {Carvalho}},\ and\ \bibinfo {author} {\bibfnamefont {A.~J.}\ \bibnamefont {Silva}},\ }\bibfield  {title} {\bibinfo {title} {Energetics and biomechanics as determining factors of swimming performance: {Updating} the state of the art},\ }\href {https://doi.org/10.1016/j.jsams.2009.01.003} {\bibfield  {journal} {\bibinfo  {journal} {Journal of Science and Medicine in Sport}\ }\textbf {\bibinfo {volume} {13}},\ \bibinfo {pages} {262} (\bibinfo {year} {2010})}\BibitemShut {NoStop}%
\bibitem [{\citenamefont {Dode}\ \emph {et~al.}(2022)\citenamefont {Dode}, \citenamefont {Carmigniani}, \citenamefont {Cohen}, \citenamefont {Clanet},\ and\ \citenamefont {Bocquet}}]{dode_wave_2022}%
  \BibitemOpen
  \bibfield  {author} {\bibinfo {author} {\bibfnamefont {A.}~\bibnamefont {Dode}}, \bibinfo {author} {\bibfnamefont {R.}~\bibnamefont {Carmigniani}}, \bibinfo {author} {\bibfnamefont {C.}~\bibnamefont {Cohen}}, \bibinfo {author} {\bibfnamefont {C.}~\bibnamefont {Clanet}},\ and\ \bibinfo {author} {\bibfnamefont {L.}~\bibnamefont {Bocquet}},\ }\bibfield  {title} {\bibinfo {title} {Wave drag during an unsteady motion},\ }\href {https://doi.org/10.1017/jfm.2022.592} {\bibfield  {journal} {\bibinfo  {journal} {Journal of Fluid Mechanics}\ }\textbf {\bibinfo {volume} {951}},\ \bibinfo {pages} {A15} (\bibinfo {year} {2022})}\BibitemShut {NoStop}%
\bibitem [{\citenamefont {Brouwer}\ \emph {et~al.}(2013)\citenamefont {Brouwer}, \citenamefont {Poel},\ and\ \citenamefont {Hofmijster}}]{brouwer_dont_2013}%
  \BibitemOpen
  \bibfield  {author} {\bibinfo {author} {\bibfnamefont {A.~J.~d.}\ \bibnamefont {Brouwer}}, \bibinfo {author} {\bibfnamefont {H.~J.~d.}\ \bibnamefont {Poel}},\ and\ \bibinfo {author} {\bibfnamefont {M.~J.}\ \bibnamefont {Hofmijster}},\ }\bibfield  {title} {\bibinfo {title} {Don’t {Rock} the {Boat}: {How} {Antiphase} {Crew} {Coordination} {Affects} {Rowing}},\ }\href {https://doi.org/10.1371/journal.pone.0054996} {\bibfield  {journal} {\bibinfo  {journal} {Plos One}\ }\textbf {\bibinfo {volume} {8}},\ \bibinfo {pages} {e54996} (\bibinfo {year} {2013})}\BibitemShut {NoStop}%
\bibitem [{\citenamefont {Weihs}(1974)}]{weihs_energetic_1974}%
  \BibitemOpen
  \bibfield  {author} {\bibinfo {author} {\bibfnamefont {D.}~\bibnamefont {Weihs}},\ }\bibfield  {title} {\bibinfo {title} {Energetic advantages of burst swimming of fish},\ }\href {https://doi.org/10.1016/0022-5193(74)90192-1} {\bibfield  {journal} {\bibinfo  {journal} {Journal of Theoretical Biology}\ }\textbf {\bibinfo {volume} {48}},\ \bibinfo {pages} {215} (\bibinfo {year} {1974})}\BibitemShut {NoStop}%
\bibitem [{\citenamefont {Weihs}(1981)}]{weihs_energetic_1981}%
  \BibitemOpen
  \bibfield  {author} {\bibinfo {author} {\bibfnamefont {D.}~\bibnamefont {Weihs}},\ }\bibfield  {title} {\bibinfo {title} {Energetic advantages of {Burst}-and-{Coast} swimming of fish at high speeds},\ }\href@noop {} {\bibfield  {journal} {\bibinfo  {journal} {Journal of Experimental Biology}\ }\textbf {\bibinfo {volume} {97}},\ \bibinfo {pages} {169} (\bibinfo {year} {1981})}\BibitemShut {NoStop}%
\bibitem [{\citenamefont {Ehrenstein}\ \emph {et~al.}(2014)\citenamefont {Ehrenstein}, \citenamefont {Marquillie},\ and\ \citenamefont {Eloy}}]{ehrenstein_skin_2014}%
  \BibitemOpen
  \bibfield  {author} {\bibinfo {author} {\bibfnamefont {U.}~\bibnamefont {Ehrenstein}}, \bibinfo {author} {\bibfnamefont {M.}~\bibnamefont {Marquillie}},\ and\ \bibinfo {author} {\bibfnamefont {C.}~\bibnamefont {Eloy}},\ }\bibfield  {title} {\bibinfo {title} {Skin friction on a flapping plate in uniform flow},\ }\href {https://doi.org/10.1098/rsta.2013.0345} {\bibfield  {journal} {\bibinfo  {journal} {Philosophical Transactions of the Royal Society A: Mathematical, Physical and Engineering Sciences}\ }\textbf {\bibinfo {volume} {372}},\ \bibinfo {pages} {20130345} (\bibinfo {year} {2014})}\BibitemShut {NoStop}%
\bibitem [{\citenamefont {Godoy-Diana}\ and\ \citenamefont {Thiria}(2018)}]{godoy-diana_diverse_2018}%
  \BibitemOpen
  \bibfield  {author} {\bibinfo {author} {\bibfnamefont {R.}~\bibnamefont {Godoy-Diana}}\ and\ \bibinfo {author} {\bibfnamefont {B.}~\bibnamefont {Thiria}},\ }\bibfield  {title} {\bibinfo {title} {On the diverse roles of fluid dynamic drag in animal swimming and flying},\ }\href {https://doi.org/10.1098/rsif.2017.0715} {\bibfield  {journal} {\bibinfo  {journal} {Journal of The Royal Society Interface}\ }\textbf {\bibinfo {volume} {15}},\ \bibinfo {pages} {20170715} (\bibinfo {year} {2018})}\BibitemShut {NoStop}%
\bibitem [{\citenamefont {Li}\ \emph {et~al.}(2021)\citenamefont {Li}, \citenamefont {Ashraf}, \citenamefont {François}, \citenamefont {Kolomenskiy}, \citenamefont {Lechenault}, \citenamefont {Godoy-Diana},\ and\ \citenamefont {Thiria}}]{li_burst-and-coast_2021}%
  \BibitemOpen
  \bibfield  {author} {\bibinfo {author} {\bibfnamefont {G.}~\bibnamefont {Li}}, \bibinfo {author} {\bibfnamefont {I.}~\bibnamefont {Ashraf}}, \bibinfo {author} {\bibfnamefont {B.}~\bibnamefont {François}}, \bibinfo {author} {\bibfnamefont {D.}~\bibnamefont {Kolomenskiy}}, \bibinfo {author} {\bibfnamefont {F.}~\bibnamefont {Lechenault}}, \bibinfo {author} {\bibfnamefont {R.}~\bibnamefont {Godoy-Diana}},\ and\ \bibinfo {author} {\bibfnamefont {B.}~\bibnamefont {Thiria}},\ }\bibfield  {title} {\bibinfo {title} {Burst-and-coast swimmers optimize gait by adapting unique intrinsic cycle},\ }\href {https://doi.org/10.1038/s42003-020-01521-z} {\bibfield  {journal} {\bibinfo  {journal} {Communications Biology}\ }\textbf {\bibinfo {volume} {4}},\ \bibinfo {pages} {1} (\bibinfo {year} {2021})}\BibitemShut {NoStop}%
\bibitem [{\citenamefont {Li}\ \emph {et~al.}(2023)\citenamefont {Li}, \citenamefont {Kolomenskiy}, \citenamefont {Liu}, \citenamefont {Godoy-Diana},\ and\ \citenamefont {Thiria}}]{li_intermittent_2023}%
  \BibitemOpen
  \bibfield  {author} {\bibinfo {author} {\bibfnamefont {G.}~\bibnamefont {Li}}, \bibinfo {author} {\bibfnamefont {D.}~\bibnamefont {Kolomenskiy}}, \bibinfo {author} {\bibfnamefont {H.}~\bibnamefont {Liu}}, \bibinfo {author} {\bibfnamefont {R.}~\bibnamefont {Godoy-Diana}},\ and\ \bibinfo {author} {\bibfnamefont {B.}~\bibnamefont {Thiria}},\ }\bibfield  {title} {\bibinfo {title} {Intermittent versus continuous swimming: {An} optimization tale},\ }\href {https://doi.org/10.1103/PhysRevFluids.8.013101} {\bibfield  {journal} {\bibinfo  {journal} {Physical Review Fluids}\ }\textbf {\bibinfo {volume} {8}},\ \bibinfo {pages} {013101} (\bibinfo {year} {2023})}\BibitemShut {NoStop}%
\bibitem [{\citenamefont {Floryan}\ \emph {et~al.}(2017{\natexlab{a}})\citenamefont {Floryan}, \citenamefont {Van~Buren},\ and\ \citenamefont {Smits}}]{floryan_forces_2017}%
  \BibitemOpen
  \bibfield  {author} {\bibinfo {author} {\bibfnamefont {D.}~\bibnamefont {Floryan}}, \bibinfo {author} {\bibfnamefont {T.}~\bibnamefont {Van~Buren}},\ and\ \bibinfo {author} {\bibfnamefont {A.~J.}\ \bibnamefont {Smits}},\ }\bibfield  {title} {\bibinfo {title} {Forces and energetics of intermittent swimming},\ }\href {https://doi.org/10.1007/s10409-017-0694-3} {\bibfield  {journal} {\bibinfo  {journal} {Acta Mechanica Sinica}\ }\textbf {\bibinfo {volume} {33}},\ \bibinfo {pages} {725} (\bibinfo {year} {2017}{\natexlab{a}})}\BibitemShut {NoStop}%
\bibitem [{\citenamefont {Akoz}\ and\ \citenamefont {Moored}(2018)}]{akoz_unsteady_2018}%
  \BibitemOpen
  \bibfield  {author} {\bibinfo {author} {\bibfnamefont {E.}~\bibnamefont {Akoz}}\ and\ \bibinfo {author} {\bibfnamefont {K.~W.}\ \bibnamefont {Moored}},\ }\bibfield  {title} {\bibinfo {title} {Unsteady propulsion by an intermittent swimming gait},\ }\href {https://doi.org/10.1017/jfm.2017.731} {\bibfield  {journal} {\bibinfo  {journal} {Journal of Fluid Mechanics}\ }\textbf {\bibinfo {volume} {834}},\ \bibinfo {pages} {149} (\bibinfo {year} {2018})}\BibitemShut {NoStop}%
\bibitem [{\citenamefont {Akoz}\ \emph {et~al.}(2021)\citenamefont {Akoz}, \citenamefont {Mivehchi},\ and\ \citenamefont {Moored}}]{akoz_intermittent_2021}%
  \BibitemOpen
  \bibfield  {author} {\bibinfo {author} {\bibfnamefont {E.}~\bibnamefont {Akoz}}, \bibinfo {author} {\bibfnamefont {A.}~\bibnamefont {Mivehchi}},\ and\ \bibinfo {author} {\bibfnamefont {K.~W.}\ \bibnamefont {Moored}},\ }\bibfield  {title} {\bibinfo {title} {Intermittent unsteady propulsion with a combined heaving and pitching foil},\ }\href {https://doi.org/10.1103/PhysRevFluids.6.043101} {\bibfield  {journal} {\bibinfo  {journal} {Physical Review Fluids}\ }\textbf {\bibinfo {volume} {6}},\ \bibinfo {pages} {043101} (\bibinfo {year} {2021})}\BibitemShut {NoStop}%
\bibitem [{\citenamefont {Akoz}\ \emph {et~al.}(2019)\citenamefont {Akoz}, \citenamefont {Han}, \citenamefont {Liu}, \citenamefont {Dong},\ and\ \citenamefont {Moored}}]{akoz_large-amplitude_2019}%
  \BibitemOpen
  \bibfield  {author} {\bibinfo {author} {\bibfnamefont {E.}~\bibnamefont {Akoz}}, \bibinfo {author} {\bibfnamefont {P.}~\bibnamefont {Han}}, \bibinfo {author} {\bibfnamefont {G.}~\bibnamefont {Liu}}, \bibinfo {author} {\bibfnamefont {H.}~\bibnamefont {Dong}},\ and\ \bibinfo {author} {\bibfnamefont {K.~W.}\ \bibnamefont {Moored}},\ }\bibfield  {title} {\bibinfo {title} {Large-{Amplitude} {Intermittent} {Swimming} in {Viscous} and {Inviscid} {Flows}},\ }\href {https://doi.org/10.2514/1.j056637} {\bibfield  {journal} {\bibinfo  {journal} {AIAA Journal}\ }\textbf {\bibinfo {volume} {57}},\ \bibinfo {pages} {3678} (\bibinfo {year} {2019})}\BibitemShut {NoStop}%
\bibitem [{\citenamefont {Gupta}\ \emph {et~al.}(2021)\citenamefont {Gupta}, \citenamefont {Thekkethil},\ and\ \citenamefont {Agrawal}}]{gupta_body-caudal_2021}%
  \BibitemOpen
  \bibfield  {author} {\bibinfo {author} {\bibfnamefont {S.}~\bibnamefont {Gupta}}, \bibinfo {author} {\bibfnamefont {N.}~\bibnamefont {Thekkethil}},\ and\ \bibinfo {author} {\bibfnamefont {A.}~\bibnamefont {Agrawal}},\ }\bibfield  {title} {\bibinfo {title} {Body-caudal fin fish-inspired self-propulsion studyon burst-and-coast and continuous swimming of a hydrofoil model},\ }\href@noop {} {\bibfield  {journal} {\bibinfo  {journal} {Physics of Fluids}\ }\textbf {\bibinfo {volume} {33}},\ \bibinfo {pages} {091905} (\bibinfo {year} {2021})}\BibitemShut {NoStop}%
\bibitem [{\citenamefont {Floryan}\ \emph {et~al.}(2017{\natexlab{b}})\citenamefont {Floryan}, \citenamefont {Van~Buren}, \citenamefont {Rowley},\ and\ \citenamefont {Smits}}]{floryan_scaling_2017}%
  \BibitemOpen
  \bibfield  {author} {\bibinfo {author} {\bibfnamefont {D.}~\bibnamefont {Floryan}}, \bibinfo {author} {\bibfnamefont {T.}~\bibnamefont {Van~Buren}}, \bibinfo {author} {\bibfnamefont {C.~W.}\ \bibnamefont {Rowley}},\ and\ \bibinfo {author} {\bibfnamefont {A.~J.}\ \bibnamefont {Smits}},\ }\bibfield  {title} {\bibinfo {title} {Scaling the propulsive performance of heaving and pitching foils},\ }\href {https://doi.org/10.1017/jfm.2017.302} {\bibfield  {journal} {\bibinfo  {journal} {Journal of Fluid Mechanics}\ }\textbf {\bibinfo {volume} {822}},\ \bibinfo {pages} {386} (\bibinfo {year} {2017}{\natexlab{b}})}\BibitemShut {NoStop}%
\bibitem [{\citenamefont {Boswell}(1971)}]{boswell_design_1971}%
  \BibitemOpen
  \bibfield  {author} {\bibinfo {author} {\bibfnamefont {R.}~\bibnamefont {Boswell}},\ }\href@noop {} {\emph {\bibinfo {title} {Design, cavitation performance, and open-water performance of a series of research skewed propellers}}},\ \bibinfo {type} {Tech. Rep.}\ \bibinfo {number} {3339}\ (\bibinfo  {institution} {Naval Ship Research and Development Center, Washington DC, USA, Department of Hydromechanics},\ \bibinfo {year} {1971})\BibitemShut {NoStop}%
\end{thebibliography}
\end{document}